\documentclass{article}
\usepackage{graphicx}
\usepackage{amsmath,amssymb}
\usepackage{graphics,color,array,calc,rotating,epsfig,psfrag}
\usepackage{hyperref}
\numberwithin{equation}{section}
\usepackage{cite}
\usepackage{bm}
\usepackage{dcolumn}
\usepackage{float}
\usepackage{subfigure}
\usepackage{tikz-cd}
\usepackage{caption}
\usepackage[utf8]{inputenc}
\usepackage{tikz}
\usepackage{amsfonts}
\usepackage[mathscr]{eucal}
\usepackage{helvet}
\def\be{\begin{equation}} \def\ee{\end{equation}}
\def\bea{\begin{eqnarray}} \def\eea{\end{eqnarray}}

\newcommand{\RN}[1]{%
  \textup{\uppercase\expandafter{\romannumeral#1}}%
}
\begin{document}
\baselineskip 18pt%
\begin{titlepage}
\vspace*{1mm}%
\hfill%
\vspace*{15mm}%
\hfill
\vbox{
    \halign{#\hfil         \cr
         IPM/P-2022/13  \cr
          } 
      }  
\vspace*{20mm}
\begin{center}
{\Large {\bf \boldmath Emergence of non-linear electrodynamic theories from\\ $T\bar{T}$-like deformations}}
 \end{center}
\vspace*{5mm}
\begin{center}
{H. Babaei-Aghbolagh$^{\dagger}$, Komeil Babaei Velni$^{*}$,
	 Davood Mahdavian Yekta$^{\ddagger}$ and H. Mohammadzadeh$^{\dagger}$
 }\\
\vspace*{0.2cm}
{\it
$^{\dagger}$Department of Physics, University of Mohaghegh Ardabili,
P.O. Box 179, Ardabil, Iran\\
$^{*}$Department of Physics, University of Guilan, P.O. Box 41335-1914, Rasht, Iran\\
$^{*}$School of Physics and School of Particles and Accelerators,\\
Institute for Research in Fundamental Sciences (IPM), P.O. Box 19395-5531, Tehran, Iran\\
$^{\ddagger}$Department of Physics, Hakim Sabzevari University, P.O. Box 397, Sabzevar, Iran\\
}

 \vspace*{0.5cm}
{E-mails: {\tt h.babaei@uma.ac.ir,  babaeivelni@guilan.ac.ir, d.mahdavian@hsu.ac.ir, mohammadzadeh@uma.ac.ir
}}
\vspace{1cm}
\end{center}

\begin{abstract}
In this letter, we investigate the deformation of the ModMax theory, as a unique Lagrangian of non-linear electrodynamics preserving both conformal and electromagnetic-duality invariance, under $T\bar{T}$-like flows. We will show that the deformed theory is the generalized non-linear Born-Infeld electrodynamics. Being inspired by the invariance under the flow equation for Born-Infeld theories, we propose another $T\bar{T}$-like operator generating the ModMax and generalized Born-Infeld non-linear electrodynamic theories from the usual Maxwell and Born-Infeld theories, respectively.
\end{abstract}

\end{titlepage}


\section{Introduction}
\label{sec1}

As the first non-linear electrodynamic (NED) generalization of Maxwell theory, the Born-Infeld (BI) model was proposed to impose an upper limit on the self-energy of charged point particles \cite{Born:1934gh}. Its Lagrangian density is a function of two independent Lorentz scalars $\mathcal{S}=-\frac14 F_{\mu\nu}F^{\mu\nu}$ and $\mathcal{P}=-\frac14F_{\mu\nu}\tilde F^{\mu\nu}$ as follows
\begin{equation}\label{BI}
 \mathcal{L}_{BI}=T\left(1-\sqrt{1-\frac{2}{T}\mathcal{S}-\frac{1}{T^2}\mathcal{P}^2}\right),
 \end{equation}
 where $\tilde F^{\mu\nu}=\frac12\, \epsilon^{\mu\nu\alpha\beta} F_{\alpha\beta}$ corresponds to the Hodge dual of the electromagnetic field strength $F_{\mu\nu}$.
 It has been shown \cite{Bialynicki-Birula:1981,Bialynicki-Birula:1992rcm} that the Lagrangian of NED theories is constrained by the equation $G\tilde{G}-F\tilde{F}=0$,  which is known as the self-duality condition \cite{Gibbons:1995ap,Gaillard:1997rt,Kuzenko:2000uh,Aschieri:2008ns,Kallosh:2011qt,Bossard:2011ij,Carrasco:2011jv,Chemissany:2011yv,Aschieri:2013nda}. The antisymmetric tensor field $G$ is defined by
\begin{equation}\label{G} G_{\mu\nu}=-2\frac{\partial{\cal L}(\mathcal{S},\mathcal{P})}{\partial{F^{\mu\nu}}},
\end{equation}
where the factor 2 takes into account because of the partial derivative with respect to the antisymmetry of ${F}_{\mu\nu}$.

A number of the NED theories such as BI, Bialynicki-Birula and Bossard-Nicolai\cite{Bossard:2011ij} satisfy the self-duality condition while the Heisenberg-Euler \cite{Heisenberg} theory does not. Though in general the action of a NED theory may not be invariant under this kind of duality, the physical objects such as the equations of motion \cite{Green:1996qg}, the energy-momentum tensor (EMT) \cite{Gaillard:1981rj,BabaeiVelni:2016qea} and the scattering amplitudes \cite{Babaei-Aghbolagh:2013hia,Garousi:2017fbe,BabaeiVelni:2019ptj} are invariant.
It has been proposed another non-linear modification of Maxwell theory that preserves the electromagnetic duality as well as conformal invariance, the so-called ModMax theory \cite{Bandos:2020jsw} which is described by
\begin{equation}
\label{LMM}
{\cal L}_{MM}=\cosh\gamma \,\mathcal{S}+\sinh \gamma\,\sqrt{\mathcal{S}^2+\mathcal{P}^2}.
\end{equation}
The unitary condition imposes the restriction $\gamma\geq 0$, which admits exact Maxwell-like plane wave solutions of arbitrary polarization \cite{Bandos:2020jsw}.
The ModMax theory could be found from dimensional reduction of a chiral 2-form ED in $D\!=\!6$ dimensions \cite{Bandos:2020hgy} obeying the self-duality condition \cite{Kosyakov}. However, the Maxwell ED is not a weak-field limit of the ModMax theory because of the conformal invariance, but it is recovered when the ModMax constant parameter $\gamma$ tends to zero \cite{Bandos:2020jsw}. More studies on this theory could be found in Refs.~\cite{Bandos:2021rqy,Kuzenko:2021cvx,Avetisyan:2021heg,Kruglov:2021bhs,BallonBordo:2020jtw,Nastase:2021uvc}.

Recently, Zamolodchikov and Smirnov proposed an exactly solvable deformation of two-dimensional (2D) quantum field theory by an irrelevant operator, known as $T \bar{T}$-deformation \cite{Zamolodchikov:2004ce,Smirnov:2016lqw}. The $T \bar{T}$ operator is constructed by the product of components of the EMT. Generalization of this operator to higher-dimensional theories with a different version of $T \bar{T}$ operator are given in Refs.~\cite{Cardy:2018sdv,Bonelli:2018kik,Taylor:2018xcy,Conti:2018jho,Babaei-Aghbolagh:2020kjg}.
One can consider a Lorentz invariant deformation of the type $S_{QFT}=S_{CFT}+S_{\lambda}$  where $S_{\lambda}=\lambda\int d^2 x \, T\bar{T}$ and
$\lambda$ is the deformation parameter. Consequently, a one parameter family of theories could be emerged by ${d S_{QFT}}/{d\lambda}=\int d^2 x \, (T\bar{T})_{\lambda}$ for finite $\lambda$ while both sides of the equation are renormalized and UV finite as well \cite{Zamolodchikov:2004ce,Cavaglia:2016oda}. In other words, a deformed Lagrangian $\mathcal{L}^{\lambda}$ can be derived by solving this $T \bar{T}$-flow equation.
For example, if we consider the $T \bar{T}$-deformation of a free massless scalar field theory, the deformed theory leads to the Nambu-Goto action where $\lambda$ is identified with the string tension $T=1/(2\pi \alpha')$ \cite{Cavaglia:2016oda}.

It would be of interest to investigate other forms of this operator in higher dimensions, especially in the light of the expected form of the deformation equation proposed by Cardy \cite{Cardy:2018sdv}, regardless of the holographic interpretations \cite{McGough:2016lol,Giveon:2017myj,Giribet:2017imm,Kraus:2018xrn,Donnelly:2018bef,Allameh:2021moy}.  For instance, in \cite{Babaei-Aghbolagh:2020kjg}, we have studied that the construction of $T \bar{T}$-deformation for some NED BI-type theories in 4D spacetime as follows
\begin{equation}
\label{TT1}
{O}_{T^2}^{\lambda}=T_{\mu\nu}T^{\mu\nu}- \frac{1}{2} {T_{\mu}}^{\mu} {T_{\nu}}^{\nu},
\end{equation}
where ${T_{\mu}}^{\mu}$ is the trace of EMT. In fact, we reproduced the expansion of 4D BI-type theories by deforming the free Lagrangian by means of the operator defined in Eq.~\eqref{TT1}.

In this letter, using the operator \eqref{TT1} and employing a simple perturbative integration technique \cite{Babaei-Aghbolagh:2020kjg}, we investigate the $T \bar{T}$-deformation of the ModMax theory.
It is shown that the emergent deformed theory is the same generalized Born-Infeld (GBI) theory given in Ref.~\cite{Bandos:2020hgy}. In addition, our main motivation is to construct a new $T\bar{T}$-like operator ${O}_{T^2}^{\gamma}$ with deformation parameter $\gamma$ under which the ModMax and GBI theories are invariant, as the BI theory which is invariant under ${O}_{T^2}^{\lambda}$.
We also show that this operator deforms the Maxwell theory to the ModMax theory with no need for additional conditions to establish the conformal and electromagnetic duality invariance as it happens in Ref.~\cite{Bandos:2020jsw}. It is proved that the GBI theory is derived from BI theory by using operator ${O}_{T^2}^{\gamma}$.

The organization of this paper is as follows: In section \ref{sec2}, we consider the deformation of the ModMax theory under $T\bar{T}$-like operator ${O}_{T^2}^{\lambda}$ by employing a perturbative approach and show that final deformed theory is equivalent to the expansion of GBI theory for small $\lambda$. We will find a new $T\bar{T}$-like operator in section \ref{sec3}, as for that the GBI theory is invariant under the flow equation $\partial {\cal L}_{\gamma BI}/{\partial \gamma} = 1/2\, O^{\gamma }_{T^2}$ where $\gamma$ is a dimensionless parameter in the NED theory. We show in section \ref{sec4} that not only the ModMax theory in invariant under the $T\bar{T}$-deformation by ${O}_{T^2}^{\gamma}$, but also can be reconstructed from Maxwell theory under the effect of this operator. Finally, in section \ref{sec5}, we provide a summary of our results and outlook.
\section{$T\bar{T}$-deformation of the ModMax theory by ${O}_{T^2}^{\lambda}$}
\label{sec2}
Starting with a free theory denoted by $\mathcal{L}_{\circ}$, one may perturbatively solve the $T\bar{T}$-flow equation ${d \mathcal{L}_\lambda}/{d\lambda}=1/8  \, (T\bar{T})$ in terms of $\lambda$ by taking into account $T_{\mu\nu}^{\lambda}$ related to the $\lambda$ order of the theory \cite{Ferko:2019oyv}. In other words, the deformation is solvable: one can make precise statements about properties of the deformed theory $\mathcal{L}_\lambda$ in terms of $\mathcal{L}_{\circ}$.
This deformation preserves integrability: if one begins with a theory $\mathcal{L}_{\circ}$ which is integrable, in the sense that the theory has infinitely many local integrals of motion, then the deformed theory $\mathcal{L}_\lambda$ at finite $\lambda$ is also integrable \cite{Cavaglia:2016oda,Bonelli:2018kik,Giveon:2017nie}.

Now, using perturbative approach, we find the interacting terms for the ModMax theory described by Eq.~\eqref{LMM}. We first compute the EMT as following
\begin{equation}
\label{EMT1}
T_{\mu\nu}=g_{\mu\nu}\, {\cal L}_{MM}+{F_{\mu}}^ \rho G_{\nu \rho},
\end{equation}
where from Eq.~\eqref{G}, the tensor $G$ is given by
\begin{equation}
\label{G1}
G_{\mu\nu}=\cosh\gamma F_{\mu\nu} +\sinh\gamma \frac{ F_{\mu\nu} \mathcal{S} + \tilde{F}_{\mu\nu} \mathcal{P}}{ \sqrt{ \mathcal{S}^2+\mathcal{P}^2 } }.
\end{equation}
Substituting Eq.~\eqref{G1} in \eqref{EMT1}, we obtain
\begin{eqnarray}
\label{TTMM}
T_{\mu\nu}= T^{Max}_{\mu\nu} \bigg(\cosh\gamma + \frac{\sinh\gamma\, \mathcal{S}}{ \sqrt{\mathcal{S}^2+\mathcal{P}^2}}\bigg),
\end{eqnarray}
where $T^{Max}_{\mu\nu}=F_{\mu\rho}{F_{\nu}}^{\rho}-1/4 \,g_{\mu\nu} F_{\alpha\beta}F^{\alpha\beta}$ is the EMT of the Maxwell theory. Due to the conformal symmetry of the ModMax theory, it would be expected that its EMT is traceless and proportional to $T^{Max}_{\mu\nu}$ \cite{Bandos:2020jsw,Sorokin:2021tge}.

Using Eq.~\eqref{TTMM} in \eqref{TT1} to construct ${O}_{T^2}^{\lambda}={O}_{0}$  at the order of $\lambda^0$, the first order $\lambda$ deformed Lagrangian is obtained as
\begin{eqnarray}\label{Lp}
{\cal L}'_{\lambda}&\!\!\!=\!\!\!&{\cal {\cal L}}_{\circ} +\frac{1}{8} \int {O}_0\,d\lambda\nonumber\\
&\!\!\!=\!\!\!&{\cal L}_{MM} + \frac{\lambda}{2} \bigg(\sinh\gamma\, \mathcal{S} +\cosh\gamma \sqrt{  \mathcal{S}^2+ \mathcal{P}^2 } \bigg)^2,
\end{eqnarray}
where we have used the convention ${O}_{0}\!=\!8 \left({\partial {\cal L}'_{\lambda}}/{\partial \lambda}\right)$ \cite{Ferko:2019oyv}.
For simplicity, we use $\mathcal{S}\!=\!\mathcal{P}\sinh(\alpha)$ so that
\begin{eqnarray}\label{def}
{\cal L}_{MM} \!=\!\mathcal{P} \sinh(\alpha + \gamma),\qquad T_{\mu\nu}= T^{Max}_{\mu\nu}\,\frac{\cosh(\alpha + \gamma)}{\cosh(\alpha)}.
\end{eqnarray}
After a straightforward calculation, Eq.~\eqref{Lp} recasts as follows
\begin{eqnarray} \label{dl1}
{\cal L}'_{\lambda}&\!\!\!=\!\!\!&\mathcal{P} \sinh(\alpha + \gamma)+\frac{1}{2} \lambda \,\mathcal{P}^2 \, \cosh^2(\alpha + \gamma) \nonumber\\
&\!\!\!=\!\!\!&{\cal L}_{MM} + \frac{1}{2} \lambda ({\cal L}_{MM} ^2 + \mathcal{P}^2).
\end{eqnarray}

Following a similar prescription, we can deform the Lagrangian to the next order of $\lambda$ as ${\cal L}{''}_{\lambda}\!=\!{\cal L}{'}_{\lambda} +{1}/{8} \int {O}_1\,d\lambda$ where ${O}_1$ is the $T\bar{T}$ operator presented by Eq.~\eqref{TT1} and it is constructed from the EMT of ${\cal L}{'}_{\lambda}$.
Iterating this method, one can deform the free theory to higher orders of $\lambda$. We have shown in \cite{Babaei-Aghbolagh:2020kjg} that if one deforms the free Maxwell theory by this method to higher orders of $\lambda$, the Maxwell BI theory will be recovered. So, we are going to investigate this implication again in this paper for the ModMax theory. For example, suppose we want to deform the ModMax theory to order $\lambda^3$, thus we need to have $G_{\mu\nu}$ and EMT at least to the order of $\lambda^2$. They are given by
\begin{eqnarray}
 G_{\mu\nu}&\!\!\!=\!\!\!&\frac{\cosh(\alpha + \gamma)\,F_{\mu \nu}+\sinh\gamma\, \tilde{F}_{\mu \nu}}{\cosh\alpha}+ \lambda \, \mathcal{P}\frac{\sinh\bigl(2 (\alpha + \gamma)\bigr)F_{\mu \nu}+2 \cosh\gamma \,\cosh(\alpha + \gamma)  \tilde{F}_{\mu \nu}}{2\cosh\alpha}\nonumber\\
&\!\!\!+\!\!\!&\lambda^2  \, \mathcal{P}^2 \frac{ \Bigl(3 \cosh\bigl(3 (\alpha \!+\! \gamma)\bigr)\!+\!\cosh(\alpha \!+\! \gamma)\Bigr)  F_{\mu \nu}\!+\!\Bigl(3 \sinh(2 \alpha \!+\! 3 \gamma)\!+\!\sinh(2 \alpha \!+\! \gamma)\!+\!2 \sinh\gamma \Bigr) \tilde{F}_{\mu \nu} }{8 \cosh\alpha},
\end{eqnarray}
and
\begin{eqnarray} \label{EMT2}
T_{\mu\nu}&\!\!\!=\!\!\!&\frac{\cosh(\alpha + \gamma)}{\cosh\alpha}\,\biggl(F_{\mu}{}^{\delta} F_{\nu \delta} + \sinh\alpha \,\mathcal{P} {g}_{\mu \nu}+\frac{1}{4} \lambda \mathcal{P}\left(4 \sinh(\alpha + \gamma)  F_{\mu}{}^{\delta} F_{\nu \delta} + \bigl(\cosh(2 \alpha + \gamma)-3 \cosh\gamma \bigr) \mathcal{P} \,\,{g}_{\mu \nu}\right)\nonumber\\
&\!\!\!+\!\!\!&\frac{1}{8} \lambda^2 \,\mathcal{P}^2 \bigl( \bigl(6 \cosh\bigl(2 (\alpha + \gamma)\bigr)-2\bigr) F_{\mu}{}^{\delta} F_{\nu \delta}  + \bigl(\sinh(3 \alpha + 2 \gamma)-2 \sinh(\alpha) - 5 \sinh(\alpha + 2 \gamma) \bigr)\mathcal{P}\, {g}_{\mu \nu}\bigr)\biggr).
\end{eqnarray}
Therefore, saving up to the third order of $\lambda$, we obtain
\begin{eqnarray}\label{MMD}
{\cal L}_{MM}^{\lambda^3} &\!\!\!=\!\!\!&{\cal L}_{MM}+\frac{1}{8}\int O_{0} \,d\lambda+\frac{1}{8} \int O_{1}\, d\lambda+\frac{1}{8} \int O_{2}\, d\lambda+\cdots \nonumber\\
&\!\!\!=\!\!\!&{\cal L}_{MM} \!+\! \frac{1}{2} \lambda \,({\cal L}_{MM}^2 \!+\! \mathcal{P}^2) \!+\! \frac{1}{2} \lambda^2 {{\cal L}_{MM}} ({\cal L}_{MM}^2 \!+\! \mathcal{P}^2)\nonumber\\
&\!\!\!+\!\!\!& \,\frac{1}{8} \lambda^3 ({\cal L}_{MM}^2 + \mathcal{P}^2) (5 {\cal L}_{MM}^2 + P^2)+{\cal O} ( \lambda^4),
\end{eqnarray}
where ${O}_{i}={O}_{T^2}^{\lambda^i}$ is the corresponding $T\bar{T}$ operator \eqref{TT1} at the order of $\lambda^i$.
We inspect that for the $i$th order of the deformed theory, the trace of the EMT satisfies the equation $T_{\mu}^{\mu}=-4\lambda \left({\partial {\cal L}^{\lambda^i}_{MM}}/{\partial \lambda}\right)$ which may be exact renormalization group equation of the $T\bar{T}$ deformed theory \cite{McGough:2016lol,Kraus:2018xrn} \footnote{It was found in Ref.~\cite{Cavaglia:2016oda} that for the special case of free scalars treated as a classical field theory, this equation holds to all orders in $\lambda$. It will be assumed that this is true for any deformed CFT.}.
We surprisingly find out that the deformed Lagrangian \eqref{MMD} is equivalent to the expansion of the following $GBI$ Lagrangian proposed in \cite{Bandos:2020hgy}
\begin{equation}\label{GBI}
{\cal L}_{\gamma BI}=\frac{1}{\lambda} \Bigg[ 1 -  \sqrt{1 -  \lambda \Bigl( 2 {\cal L}_{MM}+\lambda\mathcal{P}^2 \Bigr)} \Bigg].
\end{equation}
${\cal L}_{\gamma BI}$ is a duality-symmetric GBI NED, the combination of the ModMax and Born-Infeld theories, in which the weak-field limit yields the ModMax theory and the strong-field limit leads to the Bialynicki-Birula electrodynamics \cite{Bialynicki-Birula:1992rcm}\footnote{The Lagrangian density of GBI theory in \cite{Sorokin:2021tge} is given by
${\cal L}_{\gamma BI}=T- \sqrt{T^2 - 2T {\cal L}_{MM}-\frac{1}{16}(F_{\mu\nu}\tilde{F}^{\mu\nu})^2},$
where ${\cal L}_{MM}$ is the ModMax Lagrangian density denoted by Eq.~\eqref{LMM} and we have identified the tension $T$ with the inverse of $\lambda$ in our convention. Note also that this Lagrangian in the limit $\gamma\rightarrow 0$ goes to the BI Lagrangian given in Eq.~\eqref{BI}.}. Thus, one can write
\begin{equation}
\label{TTD}
{\cal L}_{\gamma BI} = {\cal L}_{MM}+\frac18 \int O_{T^2}^{\lambda} \,d\lambda.
\end{equation}

It has been shown \cite{Brennan:2020dkw} that the $T\bar{T}$ operator can represent a quantization method in 2D CFT. Since we have the effects of $O_{T^2}^{\lambda}$ operator in the ModMax theory, it could be interesting to use this operator in quantization of this theory and try to derive an evolution-type equation for the quantum energy spectrum. In other words, $T\bar{T}$ operator could be used to track what happens to the behavior of energies of the system as a function of the deformation parameter. In this respect, the deformation is called solvable which means some quantities in the deformed theory, such as the energy spectrum or scattering amplitude, can be computed in terms of the corresponding quantities in the un-deformed theory \cite{Smirnov:2016lqw,Cavaglia:2016oda}.

\section{New $T \bar{T}$-like deformation for constant $\gamma$ }
\label{sec3}
The connection between non-linear symmetries and $T\bar{T}$ flows might give a novel way to organize interesting low-energy effective actions in the string theory such as the BI action. In this respect, there is a natural question: are there more general classes of theories with non-linear symmetries related to the flow equations for some $T\bar{T}$-like operator? For instance, the bosonic BI action can be obtained by deforming the Maxwell theory with operator $O_{T^2}^{\lambda}$ \cite{Conti:2018jho}.

The GBI theory denoted by Eq.~\eqref{GBI} has two constant parameters $\lambda$ and $\gamma$. Thus, it is expected that one can consider two $T\bar{T}$-like operators corresponding to these couplings, i.e. $O_{T^2}^{\lambda}$ and $O_{T^2}^{\gamma}$, so that $T\bar{T}$-flow of GBI theory respecting them. From Eqs.~\eqref{GBI} and \eqref{G}, we have
\begin{eqnarray} \label{GBIG}
&&G_{\mu\nu}=\frac{\cosh\gamma\, F_{\mu \nu} \!+\! \lambda \mathcal{P} \tilde{F}_{\mu \nu}\!+\!\sinh\gamma\, (F_{\mu \nu} \mathcal{S} \!+\! \tilde{F}_{\mu \nu} \mathcal{P})(\mathcal{S}^2\!+\!\mathcal{P}^2)^{-\frac{1}{2}}}{x},\nonumber
\end{eqnarray}
where $x\!=\!\sqrt{1 -  \lambda \Bigl( 2 {\cal L}_{MM}+\lambda \mathcal{P}^2 \Bigr)}$. Substituting Eq.~\eqref{GBIG} in \eqref{EMT1}, we find
\begin{eqnarray}\label{TGBI}
T_{\mu \nu}&=&\frac{\bigl(\cosh\gamma\,\, \sqrt{\mathcal{S}^2+\mathcal{P}^2} + \sinh\gamma\,\mathcal{S}\bigr) F_{\mu}^{ \alpha} F_{\nu\alpha}}{x\sqrt{\mathcal{S}^2+\mathcal{P}^2}}  \nonumber\\
&+&\frac{\bigl(2 \cosh\gamma\, \mathcal{S} \sqrt{\mathcal{S}^2+\mathcal{P}^2}+\sinh\gamma(2 \mathcal{S}^2+\mathcal{P}^2)  \bigr) {g}_{\mu \nu}}{x\sqrt{\mathcal{S}^2+\mathcal{P}^2}}\nonumber\\
&+& \frac{\big( x-1\big) {g}_{\mu \nu}}{\lambda  x }.
\end{eqnarray}

From Eq.~\eqref{TGBI}, it is obvious that the above EMT reduces to the one that of the BI in the limit of $\gamma\rightarrow 0$. Substituting Eq.~\eqref{TGBI} in \eqref{TT1} and taking integral of the resulted equation leads to the following
\begin{eqnarray}\label{OT1}
&&\!\!\frac18 \int O^{\lambda }_{T^2} \,d\lambda \!=\!\! \int \! \frac{1\!-\!\lambda\, {\cal L}_{MM}\!-\!x}{\lambda^2 x}  \,d\lambda\nonumber\\
&&\!=\!\frac{1}{\lambda} \Bigg[ 1 -  \sqrt{1 -  \lambda \Bigl( 2 {\cal L}_{MM}+\lambda \mathcal{P}^2 \Bigr)} \Bigg]\!=\!{\cal L}_{\gamma BI }.
\end{eqnarray}
The finding of Eq.~\eqref{OT1} shows that the flow $\partial {\cal L}_{\gamma BI}/\partial \lambda$ and the $T\bar{T}$ operator have exactly the same structure and they satisfy the following equation
\begin{eqnarray} \label{flow}
\frac{\partial {\cal L}_{\gamma BI}}{\partial \lambda} = \frac18\, O^{\lambda }_{T^2},
\end{eqnarray}
which means that the GBI Lagrangian \eqref{GBI} obeys a $T\bar{T}$-flow driven by the operator $O^{\lambda }_{T^2}$. It has also been shown in Ref.~\cite{Ferko:2019oyv} that the BI theory satisfies in a similar equation.
In addition, we find out that the EMT \eqref{TGBI} fulfills $T_{\mu}^{\mu}=-4\lambda \frac{\partial {\cal L}_{\gamma BI}}{\partial \lambda}$ for all orders.

Now, we investigate the deformation of the GBI theory under a new operator $O_{T^2}^\gamma$ using the similar relationship appeared in Eq.~\eqref{flow} as the form of $\partial {\cal L}_{\gamma BI}/\partial \gamma = f(T\bar{T})$ where $f(T\bar{T})$ is a function of operator. To find an appropriate form of the new operator, we propose a function as $f \big( a \,T_{\mu\nu}T^{\mu\nu}-b\, {T_{\mu}}^{\mu} {T_{\nu}}^{\nu} \big)$ where $a$ and $b$ are constants \cite{Cardy:2018sdv}. Taking partial derivative of Eq.~\eqref{GBI} with respect to $\gamma$ yields
\begin{eqnarray}
\frac{\partial {\cal L}_{\gamma BI }}{\partial \gamma}=\frac{\cosh(\alpha + \gamma) \mathcal{P}}{x},
\end{eqnarray}
then from Eq.~\eqref{TGBI}, taking into account $\partial {\cal L}_{\gamma BI}/\partial \gamma \!=\! f(T\bar{T})$, we find the following result
\begin{eqnarray}\label{fT}
f( T \bar{T})=\frac{1}{2}  \sqrt{T_{\mu\nu}T^{\mu\nu}-\frac14 {T_{\mu}}^{\mu}{T_{\nu}}^{\nu}}.
\end{eqnarray}
Now, by introducing the new $T\bar{T}$-like operator
\begin{eqnarray}\label{TT2}
O_{T^2}^\gamma=\sqrt{T_{\mu\nu}T^{\mu\nu}-\frac14 {T_{\mu}}^{\mu}{T_{\nu}}^{\nu}},
\end{eqnarray}
we have $\partial {\cal L}_{\gamma BI }/\partial \gamma=1/2\, O_{T^2}^\gamma$. Indeed, the GBI theory is invariant under the deformation by $O_{T^2}^\gamma$ as well as by $O_{T^2}^\lambda$, i.e., ${\cal L}_{\gamma BI }$ is a solution for solvable deformations with respect to $\lambda$ and $\gamma$.

The BI theory is recovered by the deformation of the Maxwell theory with the operator $O_{T^2}^{\lambda}$ \cite{Babaei-Aghbolagh:2020kjg} and it is invariant under this deformation \cite{Ferko:2019oyv}, i.e., $ {\partial {\cal L}_{BI }}/{\partial \lambda}=1/8 \, O_{T^2}^{\lambda}$. It is of interest to know how the BI theory deforms under \eqref{TT2}. Identifying $T=\lambda^{-1}$, one can rewrite the Lagrangian density \eqref{BI} as follows
\begin{equation}
{\cal L}_{BI }  =\frac{1}{\lambda} \bigg[ 1 -  \sqrt{1 -  \lambda \bigl( 2 \mathcal{S}+\lambda \mathcal{P}^2 \bigr)} \bigg],
\end{equation}
then by defining $y=\sqrt{1-\lambda(2\mathcal{S}+\lambda \mathcal{P}^2)}$, and using Eq.~\eqref{G} in the EMT of Eq.~\eqref{EMT1} we obtain
\begin{eqnarray}
T_{\mu \nu}=\frac{1}{y} T^{Max}_{\mu \nu}+ \frac{(y+\lambda \mathcal{S} -1 ) }{\lambda\,\, y} g_{\mu \nu}.
\end{eqnarray}
Following the same procedure of the perturbative deformation introduced earlier, for the first order of $\gamma$, we have
\begin{eqnarray} \label{BIG1}
{\cal L'}_{\gamma}={\cal L}_{BI}+ \frac12  \int O_{0}^{\gamma} \,d\gamma= {\cal L}_{BI}+\gamma \frac{\sqrt{ \mathcal{S}^2+\mathcal{P}^2 }}{y },
\end{eqnarray}
where $O_{0}^{\gamma}$ refers to $O_{T^2}^{\gamma}$ at the order of ${\gamma}^0$. To find the next order of deformation, we should calculate the EMT for ${\cal L'}_{\gamma}$, that is
\begin{eqnarray} \label{EMT3}
T'_{\mu \nu}&\!\!=\!\!&\frac{1}{y} T^{Max}_{\mu \nu} + \frac{(  y+\lambda \mathcal{S} -1 ) }{\lambda \,\, y} g_{\mu \nu}\\
&\!\!+\!\!& \gamma \bigg(\!\frac{\mathcal{S}\, y^2\!+\!\lambda(\mathcal{S}^2\!+\!\mathcal{P}^2)}{y^3 \sqrt{ \mathcal{S}^2+\mathcal{P}^2 }}  T^{Max}_{\mu \nu} \!-\! \lambda  \frac{\mathcal{S} \sqrt{ \mathcal{S}^2\!+\!\mathcal{P}^2 } }{y^3} g_{\mu \nu}\bigg).\nonumber
\end{eqnarray}

Noting the relation \eqref{EMT3}, one can find the corresponding $T\bar{T}$ operator $O_{1}^{\gamma}$ and then deform theory upto the second order of $\gamma$ as
\begin{eqnarray}
{\cal L''}_{\gamma}&\!\!\!=\!\!\!&{\cal L}_{BI}\!+\! \gamma \frac{\sqrt{ \mathcal{S}^2\!+\!\mathcal{P}^2 } }{y}\!-\! \gamma^2  \frac{ (\lambda \mathcal{P}^2 \!+\! \mathcal{S}) ( \lambda \mathcal{S}\!-\!1)}{2 y^3}.
\end{eqnarray}
Continuing the perturbative procedure for arbitrary order of $\gamma$ we find out that the obtained result is actually the expansion of ${\cal L}_{\gamma BI}$ with respect to the small values of $\gamma$, i.e.,
\begin{equation} \label{TTD1}
{\cal L}_{\gamma BI}={\cal L}_{BI}+ \frac{1}{2}  \int O_{T^2}^{\gamma} \,d\gamma .
\end{equation}

\section{ModMax theory as $T\bar{T}$-deformation of the Maxwell theory by $O_{T^2}^{\gamma}$}
\label{sec4}
In the following, we consider the deformation of the ModMax and Maxwell theories by using the new $T\bar{T}$-like operator $O_{T^2}^{\gamma}$. In the first case, we use the EMT \eqref{TTMM} to construct the operator $O_{T^2}^{\gamma}$ according to the definition \eqref{TT2}. Then, taking the partial derivative of the Lagrangian density of the ModMax theory given by \eqref{LMM} with respect to $\gamma$, we obtain
\begin{equation}
\frac{\partial {\cal L}_{MM}}{\partial \gamma}=\frac12\, O_{T^2}^{\gamma},
 \end{equation}
which states that the ModMax theory is a solution of the flow equation, just like BI with $O_{T^2}^{\lambda}$ or GBI under $O_{T^2}^{\gamma}$ deformations. In other words, the ModMax theory is the only conformal duality-invariant NED theory. It has been shown that $\mathcal{L}_{MM}$ and $O_{T^2}^{\gamma}$ make up a pair of canonically conjugate variables which can be regarded as local coordinates of a 2D symplectic manifold \cite{Kosyakov:2022klk}.

For the latter case, substituting the EMT of the Maxwell theory in \eqref{TT2}, one can derive the first order operator as $O_{0}^{\gamma}=2\sqrt{\mathcal{S}^2+\mathcal{P}^2}$. Then, the deformed Lagrangian upto the first order according to ${\cal L'}_{\gamma}={\cal L}_{Max}+ 1/2 \int O_{0}^{\gamma} \,d\gamma$ becomes
\begin{eqnarray}
{\cal L'}_{\gamma}=\mathcal{S}+\gamma   \sqrt{\mathcal{S}^2+\mathcal{P}^2},
\end{eqnarray}
where ${\cal L}_{Max}=\mathcal{S}$. For the next order, we should find the EMT corresponding to $\cal L'_{\gamma}$, then from Eqs.~\eqref{G} and \eqref{EMT1}, we obtain
\begin{eqnarray}
T'_{\mu\nu}= T^{Max}_{\mu \nu}\bigl(1 +\gamma \frac{  \mathcal{S}}{\sqrt{\mathcal{S}^2+\mathcal{P}^2}}\bigr).
\end{eqnarray}
Constructing the operator $O_{1}^{\gamma}$ with $T'_{\mu\nu}$, the deformed theory up to the second order is following
\begin{eqnarray}
{\cal L}^{''}_{\gamma}&=&\mathcal{S}+\gamma   \sqrt{\mathcal{S}^2+\mathcal{P}^2} +\frac{1}{2}  \gamma^2 \mathcal{S}.
\end{eqnarray}

One can proceed step by step to deform the Maxwell theory to higher orders of $\gamma$. We have done this calculation up to the 7th order as follows
\begin{eqnarray}
{\cal L}^{(7)}_{\gamma}&\!\!=\!\!&\mathcal{S} + \gamma\sqrt{\mathcal{S}^2+\mathcal{P}^2} + \tfrac{1}{2} \gamma^2 \mathcal{S}  + \tfrac{1}{6} \gamma^3 \sqrt{\mathcal{S}^2+\mathcal{P}^2}+ \tfrac{1}{24} \gamma^4 \mathcal{S}  \nonumber\\
&\!\!+\!\!& \! \tfrac{1}{120} \gamma^5 \sqrt{\mathcal{S}^2\!+\!\mathcal{P}^2} \!+\! \tfrac{1}{720} \gamma^6 \mathcal{S} \! + \!\tfrac{1}{5040} \gamma^7 \sqrt{\mathcal{S}^2\!+\!\mathcal{P}^2}.
\end{eqnarray}
This expression is exactly equal to the expansion of the ModMax theory given in Eq.~\eqref{LMM} with respect to the small values of $\gamma$, thus similar to the Eqs.~\eqref{TTD} and \eqref{TTD1} one can write
\begin{eqnarray} \label{TTD2}
 {\cal L}_{MM}={\cal L}_{Max}+ \frac{1}{2}  \int O_{T^2}^{\gamma} \,d\gamma.
\end{eqnarray}
\section{Conclusion and outlook}
\label{sec5}
We have studied diverse deformations of the NED extensions of the Maxwell theory under $T\bar{T}$-like operators, in particular, the ModMax theory as a unique conformal and duality invariant modification of the Maxwell theory. We shown that the electrodynamic BI action and the GBI theory can be respectively constructed by deforming the Maxwell and the ModMax Lagrangian density by operator $ {O}_{T^2}^{\lambda}$ where $\lambda$ is a typical deformation parameter. On the other hand, it was found that both of the BI and GBI theories are invariant under the flow equation with respect to ${\lambda}$.
In addition, since the ModMax theory has a non-negative constant $\gamma$, which could be regarded as a deformation parameter, we proposed a $T\bar{T}$-like operator $O_{T^2}^{\gamma}$ in Eq.~\eqref{TT2} such that the GBI theory is invariant under the flow equation of this parameter.

In order to better understand the sequence flow of these operators, we have also provided a schematic chart for the application of two operators on the Maxwell theory and the corresponding emergent deformed theories in Fig~\ref{fig}. It is worth mentioning that there is a main difference between these operators in the number of higher derivative terms that produce through deformation process and may be important in investigating the integrability and quantization of the emergent theories. The operator ${O}_{T^2}^{\lambda}$ increases the derivatives of the theory which means the enhancement of the levels of energy where the traceless property of the EMT is not necessarily retained, while, in contrast, the operator $O_{T^2}^{\gamma}$ generates terms of the order of the Maxwell action and keeps the traceless property. However, the electromagnetic duality is respected under the deformation of two operators.

\begin{center}
\begin{tikzcd}
{\cal L}_{Max} \arrow[r, blue, "{O}_{T^2}^{\lambda}" blue] \arrow[d,red,"O_{T^2}^{\gamma}" red]
&|[blue]| {\cal L}_{BI} \arrow[d,red, "O_{T^2}^{\gamma}" red] \\
|[red]|{\cal L}_{MM} \arrow[r, blue, "{O}_{T^2}^{\lambda}" blue]
&|[red!50!blue]|  {\cal L}_{\gamma BI}
\end{tikzcd}
\captionof{figure}{Sequence deformations of the Maxwell theory under ${O}_{T^2}^{\lambda}$ and $O_{T^2}^{\gamma}$.}\label{fig}
\end{center}

In the context of $T\bar{T}$-deformation that shifts a ground state theory to a class of higher excited-state theories, one can take a closer look at $T\bar{T}$ flow in higher dimensional gauge theories. For example, it has been shown \cite{Bandos:2020hgy} that there is a generic 6D chiral 2-form ED which in
the weak-field limit is related to the dimensional reduction to the ModMax theory. So, it would be of interest to examine a kind of deformation that produces some conformal and chiral theories in $D=6$ dimensions. From 4D point of view, the structure of the corresponding 6D operators may also be related to $O_{T^2}^{\gamma}$ and ${O}_{T^2}^{\lambda}$ operators studied in this letter.

In addition, the supersymmetric extension of the Born-Infeld Lagrangian has been already studied in Ref.~\cite{Bagger:1996wp}. The $T\bar{T}$-deformation of the $N=1$ supersymmetric BI theory with operator $O_{T^2}^{\lambda}$, was considered for the first time in Ref.~\cite{Ferko:2019oyv}. Also, we have studied the deformation of $N=2$ supersymmetric BI theory by this operator in \cite{Babaei-Aghbolagh:2020kjg}. The $T\bar{T}$-deformation of the $N=1$ supersymmetric ModMax theory with the operator $O_{T^2}^{\lambda}$ , has been recently studied in Ref.~ \cite{Ferko:2022iru}. Extension of the $N=1$ and $N=2$ supersymmetric ModMax theory with operator $O_{T^2}^{\gamma}$ can be investigated as a future work.
Since the ModMax theory is unitary for non-negative values of $\gamma$ \cite{Sorokin:2021tge}, it is believed that the $T\bar{T}$-deformation can be used to quantize a theory from the corresponding ground state. Therefore, providing a quantum representation for the states corresponding to the different NED theories would be important.

In general, the electromagnetic duality in the pure NED theories is described by the $SO(2)$ symmetry group \cite{Bialynicki-Birula:1981,Bialynicki-Birula:1992rcm,Gibbons:1995ap,Gaillard:1997rt}. However, one can enhance this symmetry in the NED theories to $SL(2,R)$ group by adding some axion-dilaton fields to the Lagrangian density \cite{Gaillard:1997rt}. In the case of ModMax theory as a conformal NED theory, we propose that
\begin{eqnarray}\label{sl2}
\hat{\mathcal{L}}_{MM}=\cosh \gamma \,e^{-\phi_0} \mathcal{S}+\sinh \gamma\, \sqrt{e^{-2\phi_0}\,(\mathcal{S}^2+\mathcal{P}^2)}-C_{0} \mathcal{P},
\end{eqnarray}
where $C_0$ and $\phi_0$ are the axion and dilaton fields, respectively. We postpone the details of this $SL(2,R)$-invariant ModMax theory as future work in progress \cite{Hosein:2022}.
\section*{Acknowledgment}
We are very grateful to D. Sorokin for his kindly interest in this work and fruitful
discussion. HBA is specially appreciate M.M. Sheikh-jabbari for his useful comments at the early stages of the work. The authors would also like to thank M. R. Garousi, A. Ghodsi and H. R. Afshar for valuable discussions.

\if{}
\bibliographystyle{abe}
\bibliography{references}{}
\fi

\providecommand{\href}[2]{#2}\begingroup\raggedright\endgroup

\end{document}